\documentclass[review]{elsarticle}

\usepackage{hyperref}

\journal{QIP}

\usepackage{color}
\newcommand{\tr}[1]{\,\mathrm{Tr}\left\lbrace  #1 \right\rbrace}
\newcommand{\ket}[1]{\left\vert#1\right\rangle}
\newcommand{\bra}[1]{\left\langle#1\right\vert}

\begin{document}

\begin{frontmatter}

\title{Multipartite quantum and classical correlations in symmetric $n$-qubit mixed states}

\author{Gian Luca Giorgi}
\address{Istituto Nazionale di Ricerca Metrologica, Strada delle Cacce 91, I-10135 Torino, Italy\\ and  \\ Instituto de F\'isica Interdisciplinar y Sistemas Complejos IFISC (CSIC-UIB), Campus Universitat Illes Balears, E-07122 Palma de Mallorca, Spain \\
gianluca@ifisc.uib-csic.es}

\author{Steve Campbell}
\address{Centre for Theoretical Atomic, Molecular and Optical Physics\\ School of Mathematics and Physics, Queen's University\\
Belfast BT7 1NN, United Kingdom\\
steven.campbell@qub.ac.uk}

\begin{abstract}
We discuss how to calculate genuine multipartite quantum and classical correlations in symmetric, spatially invariant, mixed $n$-qubit density matrices. We show that the existence of symmetries greatly reduces the amount of free parameters  to be optimized in order to find the optimal measurement that minimizes the conditional entropy in the discord calculation. We apply this approach to the states exhibited dynamically during a thermodynamic protocol to extract maximum work. We also apply the symmetry criterion to a wide class of physically relevant cases of spatially homogeneous noise over multipartite entangled states. Exploiting symmetries we are able to calculate the nonlocal and genuine quantum features of these states and note some interesting properties.
\end{abstract}


\end{frontmatter}


\section{Introduction}	

How quantum systems can be correlated is typically discussed within three categories: non-local, non-separable (entangled), or non-classical (discordant). Their formal differences aside, one common aspect of them all is the difficulty in extending any definition to mixed multipartite systems. Quantum discord was originally defined as the difference between two quantum analogues of the classical mutual information~\cite{Ollivier,Henderson}. The interest around such a quantumness quantifier is justified considering  the fact that there are examples in mixed-state quantum computation where it appears clear that entanglement in not the only meaningful indicator~\cite{info-no-ent1,info-no-ent2,info-no-ent3}. Even if the role of discord in quantum computation has not yet been fully clarified, there are many contexts where its use has been productive~\cite{gu,gu2,gu3,gu4}.

The original definition of discord, being based on the separation between system and apparatus~\cite{Ollivier}, applies naturally to bipartite systems. As in the case of entanglement~\cite{horod}, the extension to the multipartite scenario is not trivial. A series of postulates any good measure of multipartite correlations should obey was given in Ref.~\cite{grudka} by Bennett \textit{et al.}. Three main generalizations of quantum discord to the multipartite case have been proposed: Rulli and Sarandy introduced  the so called global discord (GD) which is a natural extension of a symmetrized version of the QD  and is based
on a collective measurement~\cite{Rulli},  Modi  and co-workers proposed a unified view of correlations that applies  in both the bipartite and in the multipartite scenarios based on
the use of the relative entropy to quantify the ``distance'' between states~\cite{modi}, in Ref.~\cite{genuine}, relative entropy was also employed to give a measure of genuine total, classical, and quantum correlations.
Genuine correlations were defined as the amount of correlation that cannot be accounted for considering any of the possible subsystems.  It was explicitly described how to implement the general definition to the case of three-qubit pure states. However, due to consistency problems, it is not clear to what  states that definition can be easily applied.

The calculation of quantum discord is in general a very hard task. Also in the bipartite scenario, exact analytical results are known only in a few special cases. The case of two qubits was explored in Refs. \cite{luo,girolami,shi}, with some further bounds given in~\cite{bounds1,bounds2,bounds3}, the Gaussian discord was presented in \cite{paris,adesso}, while the analysis of quantum correlations in high-dimensional states was performed by Chitambar in Ref. \cite{chitambar}.

 A major obstacle is represented by the fact that a minimization over a complete set of  positive-operator valued measures (POVMs) must be performed. This problem is generally overcome observing that, in common cases, orthogonal-projective measurements give a fairly tight upper bound on discord~\cite{epl}.
The problem of quantifying discord becomes, obviously, even more complicated in the multipartite scenario, where too many parameters need to be taken into account. However, the complexity of the problem is dramatically reduced in the presence of states that enjoy some symmetry property. Thus, in this work we make a significant step in the quantitative analysis of multipartite correlations, taking inspiration from previous studies by focusing on some classes of physically relevant states.

The first scenario we consider, where symmetric states play  a fundamental role, can be found in quantum thermodynamics.
Recent works have started investigating the possibility of exploiting the degree of quantumness of a system in order to achieve advantage in thermodynamic processes. More specifically, the role of entanglement was studied in Refs.~\cite{alicki,hovna} in connection to the maximal work extraction problem in finite quantum systems under cyclic transformations~\cite{Allahverdyan}. Considering  the same physical process, in Ref.~\cite{work}, we analyzed the behavior of multipartite quantum discord, using both the definition of global discord~\cite{Rulli,campbellGD} and the measure of genuinely multipartite correlations~\cite{genuine}. The protocol consists of a series of swap operations between different eigenstates of a density matrix that is diagonal with respect to the Hamiltonian basis both at the beginning and at the end of the cycle.  As the state conserves its spatial invariance during the cycle, it is possible to apply symmetry considerations.

An equally important line of inquiry has been the dynamics of quantum correlations in open systems. In particular, understanding how quantum correlations behave under adverse effects is interesting from both a pragmatic viewpoint regarding the utility of such correlations, and in understanding the differences and similarities arising when studying different types of quantum correlations. Only recently has the study of multipartite non-local~\cite{acin,campbellNL} and non-classical~\cite{maziero,fanchini} correlations in such adverse situations been studied since the calculation of these quantities is typically extremely difficult, in fact, we remark that this observation was made rigorous recently in Ref.~\cite{yichen} where the calculation of discord was proven to be NP-complete. We show for some widely applicable noisy processes that we can again efficiently calculate both the non-local and genuinely quantum correlations for a class of symmetric states. 

In this work, motivated by these physically relevant  examples, we address the question of quantitatively studying genuine correlations, global discord, and multi-partite nonlocality for spatially symmetric $n$-qubit states, where we achieve analytic expressions, and further note some interesting differences between the figures of merit under the action of noisy channels. Some complementary analyses can also be found in Refs.~\cite{beggi1,beggi2} where multipartite correlations are quantitatively explored.

\section{Measures and indicators of multipartite correlations}
\subsection{Genuine multipartite correlations}
\label{genuinediscord}
In this section we briefly review the definition of genuine quantum and classical correlations given in Ref.~\cite{genuine}. Given an $n$-partite density matrix $\varrho$ and its reduced states $\varrho_j$ ($j=1,\dots,n$),  genuine  correlations can be defined starting from  the generalization of mutual information to $n$ parties $T(\varrho)$ as a measure of total correlations: 
\begin{equation}
T(\varrho)=\sum_{j=1}^n S(\varrho_j)-S(\varrho),
\end{equation}
where $S(.)$ is the von Neumann entropy.
Genuine correlations represent the amount of correlations that cannot be accounted for considering any of the possible reduced subsystems: 
an $n$-partite state has genuine $n$-partite correlations if it is non-product in every bipartite cut (this definition is in agreement with the general criteria given in Ref.~\cite{grudka}). According to this criterion, genuine total correlations $T^{(n)}(\varrho)$ coincide with the distance, measured through the relative entropy, between $\varrho$ and the closest state with no $n$-partite correlations, that is, the closest state which is product at least along a bipartite cut \cite{modi}. This definition implies that $T^{(n)}$ coincides with the minimum bipartite mutual information present in the system. Then, we can obtain quantum (${\cal D}^{(n)}$) and classical (${\cal J}^{(n)}$) genuine correlations just applying the usual definition of quantum discord and classical correlations for bipartite systems, where the bipartition is defined along the optimal cut along which $T^{(n)}$ is calculated.

Trying to extend such a relative entropy-based criterion to any level of separability, it turns out that a consistent set of definitions can not always be given. In fact, even in the simple tripartite case, it can happen that the sum of tripartite and bipartite quantum (or classical) correlations exceeds its total value. This is due to the fact that the different sub-parts are in general different to each other, which causes the different combinations of correlations not to sum as desired. This problem is obviously not present if the state under investigation is spatially symmetric, which will be the case for the states we discuss in this work.

\subsection{Global discord}
The global discord (GD) is a multipartite extension of the original definition of the bipartite discord when collective measurements are applied~\cite{Rulli}. It is defined
\begin{equation}
\label{GQD}
\mathcal{G}_n(\rho)=\min_{\{\hat\Pi^p\}}\left\{S\left(\rho || \hat\Pi(\rho)\right)-\sum_{j=1}^n S\left(\rho_{j}||\hat\Pi_j(\rho_{j})\right)\right\},
\end{equation}
with $\rho$ the density matrix for the total state, $\rho_j=\mathrm{Tr}_{i\neq	j}\left[\rho \right]$ the reduced state of qubit $j$, $\hat\Pi_j(\rho_{j})=\sum_{l}\hat\Pi_{j}^{l}\rho_{j}\hat\Pi_{j}^{l}$, and $\hat\Pi(\rho)=\sum_k \hat\Pi^k \rho \hat\Pi^k$, where $\hat \Pi^k=\otimes^n_{l=1}\hat \Pi^{k_l}_{l}$, and $k$ stands for the string of indices $(k_1\dots k_n)$. Here, the symbol $\hat\Pi$ denotes standard von Neumann projectors, upon which the minimum has to be found, contrarily to what happens for the standard quantum discord, where POVMs are invoked~\cite{Rulli}.

 Differently to the measure presented in the previous section, the GD is not sensitive to genuine $n$-partite correlations, i.e. states exhibit quantum correlations among some of its subsystems but are nevertheless separable over some bipartition. However, it is a reliable indicator of quantumness in a given state, a non-zero GD guarantees that at least two of the subsystems exhibit quantum correlations.

In general this quantity is difficult to calculate again due to the required minimization in Eq.~(\ref{GQD}), however symmetries can help simplify the calculation. An alternative formulation that reduces the computational effort was given in Ref.~\cite{campbellGD}, where the multi-qubit projections were expressed in terms of local multi-qubit rotations, $\hat{R}_i(\theta_i,\phi_i)=\cos\theta_i\hat1\!\!1+i\sin\theta_i\cos\varphi_i\hat\sigma_y+i\sin\theta_i\sin\varphi_i\hat\sigma_x$ applied to the separable eigenstates of $\otimes_{i=1}^n\hat\sigma_z$. The minimization of Eq.~(\ref{GQD}) of an arbitrary state is then a minimization over the $2n$ angles associated with the rotations. When the state is symmetric, as we shall show, it is possible to greatly reduce the number of parameters to minimize over.

\subsection{Multipartite non-locality}
The final indicator we will assess is the multipartite non-locality based on the violation of a Bell-type inequality. There exist a wide variety of such inequalities, each with its own merits, we will restrict ourselves to extension of the tripartite Svetlichy inequality~\cite{svet} to $n$-partite systems given by Collins {\it et al}~\cite{gisin}, wherein an iterative means to to construct multipartite Bell inequalities for dichotomic observables was given. Taking $o_j$ and $O_j$ as the two outcomes of a measurement performed over one of them and setting $m_1=o_1$ and $M_1=O_1$, the polynomials are given by
 \begin{eqnarray}
 \label{theory}
m_n&=\frac{1}{2}m_{n-1}(o_n+O_n)+\frac{1}{2}M_{n-1}(o_n-O_n),\\
M_n&=\frac{1}{2}M_{n-1}(o_n+O_n)+\frac{1}{2}m_{n-1}(O_n-o_n).
 \end{eqnarray}
We can then define the generalized Svetlichny polynomials 
\begin{eqnarray}
\label{SI}
&{\cal N}_n=m_n~~~(n~\mathrm{even})\\
&{\cal N}_n=(m_n+M_n)/2~~~(n~\mathrm{odd})\nonumber
\end{eqnarray}
The bound imposed on ${\cal N}_n$ by local hidden-variable models is $1$, while quantum mechanically an $n$-qubit GHZ state achieves the maximum value of $\sqrt{2^{n-1}}$ and $\sqrt{2^{n-2}}$ for an even and odd number of qubits, respectively. These inequalities are particularly useful as they detect genuine multipartite non-locality similar to the genuine correlations outlined in Sec.~\ref{genuinediscord}. Additionally, depending on the degree of violation we can determine if there exists a $k$-separable hidden variable model to reproduce the correlations in a given state. While formally there is no way to `quantify' the non-locality in a state, an intuitive picture can be obtained by considering a states resilience to noise before it no longer violates the inequality. 

\section{Genuine correlations of symmetric states}\label{sec:gen}
Given a symmetric $n$-partite state,  genuine quantum (classical) correlations  are, unambiguously, the quantum (classical) part of the minimum bipartite correlation in the state. Following the same criterion also $m$-partite correlations ($m<n$) can be calculated.  Despite the existence of a clear definition, as calculating discord requires a minimization procedure, it is in general hard to find analytical results. However, the presence of symmetries can greatly simplify the calculation.

Our goal is to calculate any $\{n-k :k\}$ discord of an $n$-partite state, that is, the discord between the first $n-k$ and  last $k$ sub-parts of the state (or any other spatial combination). As explained in the introduction, quantum discord is obtained minimizing the conditional entropy over a complete set of POVMs. On the other hand, limiting the calculation to orthogonal projectors, as largely discussed in the literature, gives a fairly tight upper bound  \cite{epl} and is the approach we will utilize throughout this work. 

The elements of a complete set of orthogonal projectors representing the measurement process on $k$ sub-parts can be generically written as $B_i=|\psi_i\rangle \langle\psi_i|$, with
\begin{equation}
|\psi_i\rangle=\sum_i \alpha_i |i  \rangle,
\end{equation}
where $|i  \rangle$ is any of the states of the $k$-partite  basis. By definition of bipartite discord,
\begin{equation}
{\cal D}_{n-k :k}=S(\varrho_{\{k\}})-S(\varrho)+\sum_i b_i S(\langle\psi_i|\varrho|\psi_i\rangle/b_i),\label{disc}
\end{equation}
where $\varrho_{\{k\}}$ is the reduced density matrix over the $k$ sub-parts of the state, $b_i=\tr{\langle\psi_i|\varrho|\psi_i\rangle}$, and where the measurement process minimizing the conditional entropy has been already found. Then, as shown in Ref.~\cite{chiribella} for any extremal POVM, the symmetry of the states is reflected in a symmetry of the optimal measurements, which  can be assumed to be covariant with respect to the same symmetry group.

Let us assume the existence of a symmetry operator acting on the $k$ qubits to be measured   under the action of which the total density matrix $\varrho$ is left unchanged:
\begin{equation}
\varrho=U_{\{k\}}^{\dag} \varrho \, U_{\{k\}}.\label{transl}
\end{equation}
If we put Eq.~(\ref{transl}) into Eq.~(\ref{disc}) we get
\begin{equation}
{\cal D}_{n-k :k}=S(\varrho_{\{k\}})-S(\varrho)+\sum_i b_i S(\langle\tilde\psi_i|\varrho|\tilde\psi_i\rangle/b_i)\label{disc2}
\end{equation}
where $|\tilde\psi_i\rangle=U_{\{k\}}|\psi_i\rangle$.
Notice that the coefficients $b_i$ are not modified. As a consequence, the comparison between 
  Eq.~(\ref{disc}) to  Eq.~(\ref{disc2}) leads us to the conclusion  that the optimal basis $\{|\psi_i\rangle\}$ is made  by eigenstates of $U_{\{k\}}$. 
     Indeed, for any $i$, that is, term by term, $\langle \tilde\psi_i | \varrho| \tilde\psi_i\rangle=\langle \psi_i | \varrho| \psi_i\rangle$. Writing $| \tilde\psi_i\rangle=\sum_k c_{ik}| \psi_k\rangle$ we arrive to $\sum_{k,k^{\prime}}c^*_{ik}c_{ik^{\prime}}\langle \psi_k | \varrho| \psi_{k^{\prime}}\rangle=\langle \psi_i | \varrho| \psi_i\rangle$, which always admits the solution $c_{ik}=\delta_{ik}e^{i\phi_i}$. Then, $U_{\{k\}}|\psi_i\rangle=e^{i\phi_i}|\psi_i\rangle$.

\section{Multipartite correlations in a thermodynamic process}
\label{thermo}
\subsection{Genuine correlations}
A clear example of the application of this criterion is given considering the maximal work extraction protocol studied in Ref. \cite{work}. In particular we consider the symmetric $n$-qubit state
\begin{eqnarray}
\varrho&=&\frac{p_0^n+p_1^n}{2}(|00\dots 0\rangle\langle 00\dots 0|+|11\dots 1\rangle\langle 11\dots 1|)\nonumber\\&+&
\frac{p_0^n-p_1^n}{2}(|00\dots 0\rangle\langle 11\dots 1|+|11\dots 1\rangle\langle 00\dots 0|)+
\sum_{j=1}^{n-1}p_0^j p_1^{n-j}{\cal I}_j,
\label{rho}
\end{eqnarray}
where ${\cal I}_j$ is the identity operator in the $j$-excitation subspace, and in general $0\leq p_0 \leq1$ with $p_1 = 1-p_0 $. This state can be obtained starting from $\left(p_{0}|0\rangle\langle 0|+p_{1}|1\rangle\langle 1|\right)^{\otimes n}$ by coherently  mixing $|0\rangle^{\otimes n}$ to $|1\rangle^{\otimes n}$ to the end of reordering these eigenvalues of the state at the end of the cycle.
 
For this density matrix, two symmetry operators can be identified: the translation operator $\mathcal{T}$ (here, for the sake of clarity, with the word translation we mean any spatial manipulation of the state), which embodies the spatial invariance of the state, and parity-related operator (the parity $\mathcal{P}$ tells us if in the state there is an even or odd  number of $0$s). Let us consider the form of the symmetric eigenstates explicitly for different values of $k$ (starting from $k=2$, as we need a multipartite basis for the measurement).

\begin{itemize}
\item  $ k=2$
The symmetries are $\mathcal{T}$ and the parity $\mathcal{P}$. The family of common eigenvectors of these operators is
\begin{eqnarray}
|\psi_1\rangle&=&\cos\theta |00\rangle + \sin\theta |11\rangle\nonumber\\
|\psi_2\rangle&=&-\sin\theta |00\rangle + \cos\theta |11\rangle\nonumber\\
|\psi_3\rangle&=&\frac{|01\rangle+|10\rangle}{\sqrt{2}}\nonumber\\
|\psi_4\rangle&=&\frac{|01\rangle-|10\rangle}{\sqrt{2}}
\end{eqnarray}
So, the minimization necessary  to calculate discord is reduced to finding the optimal value of a single parameter $\theta$ (which is found to be $\pi/4$). Then,  the Bell basis is optimal.

\item $ k=3$

In this case, any translation operator acting on the system  commutes with $\varrho$, while the parity does not. Nevertheless, a symmetry can be found  by multiplying $\mathcal{P}$ by an operator, acting on the three qubits, such that $|0\rangle \to |0\rangle $ and $|1\rangle \to e^{i \pi/3} |1\rangle $ (let us call this symmetry operator $\mathcal{P}^{(3)}$). Eigenstates of $\mathcal{P}^{(3)}$ invariant under translation are 
$|000\rangle$ and $|111\rangle$ with eigenvalue $+1$, $|W\rangle=(|001\rangle+|010\rangle+|100\rangle)/\sqrt{3}$ and two orthogonal combinations more $|W^\prime\rangle$ and  $|W^{\prime\prime}\rangle$ (one of them could be, for instance, $(|001\rangle+e^{i 2 \pi/3}|010\rangle+e^{-i 2 \pi/3}|100\rangle)/\sqrt{3}$) with eigenvalue $-e^{i\pi/3}$, and  $|\bar W\rangle=(|011\rangle+|101\rangle+|110\rangle)/\sqrt{3}$ together with two orthogonal combinations $|\bar W^{\prime}\rangle$ and  $|\bar W^{\prime\prime}\rangle$
 with eigenvalue $e^{2i\pi/3}$. The optimal basis is given by 
\begin{eqnarray}
|\psi_1\rangle&=&\cos\theta |000\rangle + \sin\theta |111\rangle\nonumber\\
|\psi_2\rangle&=&-\sin\theta |000\rangle + \cos\theta |111\rangle\nonumber\\
|\psi_3\rangle&=&|W\rangle\nonumber\\
|\psi_4\rangle&=&|W^{\prime}\rangle\nonumber\\
|\psi_5\rangle&=&|W^{\prime\prime}\rangle\nonumber\\
|\psi_6\rangle&=&|\bar W\rangle\nonumber\\
|\psi_7\rangle&=&|\bar W^{\prime}\rangle\nonumber\\
|\psi_8\rangle&=&|\bar W^{\prime\prime}\rangle\nonumber\\
\end{eqnarray}
Actually, given that when evaluating $\langle\psi_i|   \varrho |\psi_i\rangle$ phases do not matter, we can safely use $|\psi_1\rangle,\;|\psi_2\rangle$, three times $|W\rangle$, and three times $|\bar W\rangle$. As in the $k=2$ case,  the optimal value of $\theta$ is found to be $\pi/4$. The optimal basis  then consists of maximally entangled states.

\item $ k>3$
The generalization to any $k$ is straightforward. A symmetry operator $\mathcal{P}^{(k)}$ always exists and is the product of the parity $\mathcal{P}$ by a second operator  such that $|0\rangle \to |0\rangle $ and $|1\rangle \to e^{i \pi/k} |1\rangle $ if $k$ is odd and $|1\rangle \to e^{2 i \pi/k} |1\rangle $ if $k$ is even. The optimal set of measurements is represented by $|GHZ_{\pm}\rangle=(|0\rangle^{\otimes k}\pm |1\rangle^{\otimes k})/\sqrt{k}$ and by the generalized $|W\rangle$ state (Fourier mode) in any of the $j$-excitation subspaces ($j=1,\dots,k-1$) with multiplicity $ {k}\choose{j} $. 
\end{itemize}

So, despite the multipartite nature of the state a number of operations of size $k$ is sufficient to calculate any bipartite discord between many-qubit parts. In order to get the true $n$-partite discord, a quantitative comparison among  all the possible cuts is finally required. In Fig.~\ref{fig1} we show the behavior of ${\cal D}^{(n)}$ as a function of the state parameter $p_0$ for different values of $n$. Any of the lines plotted has been obtained taking the minimum over all the possible $\{n-k :k\}$ partitions.

\subsection{Global discord}
\begin{figure}[t]
\begin{center}
(a) \hskip5cm (b)
\includegraphics[width=5cm]{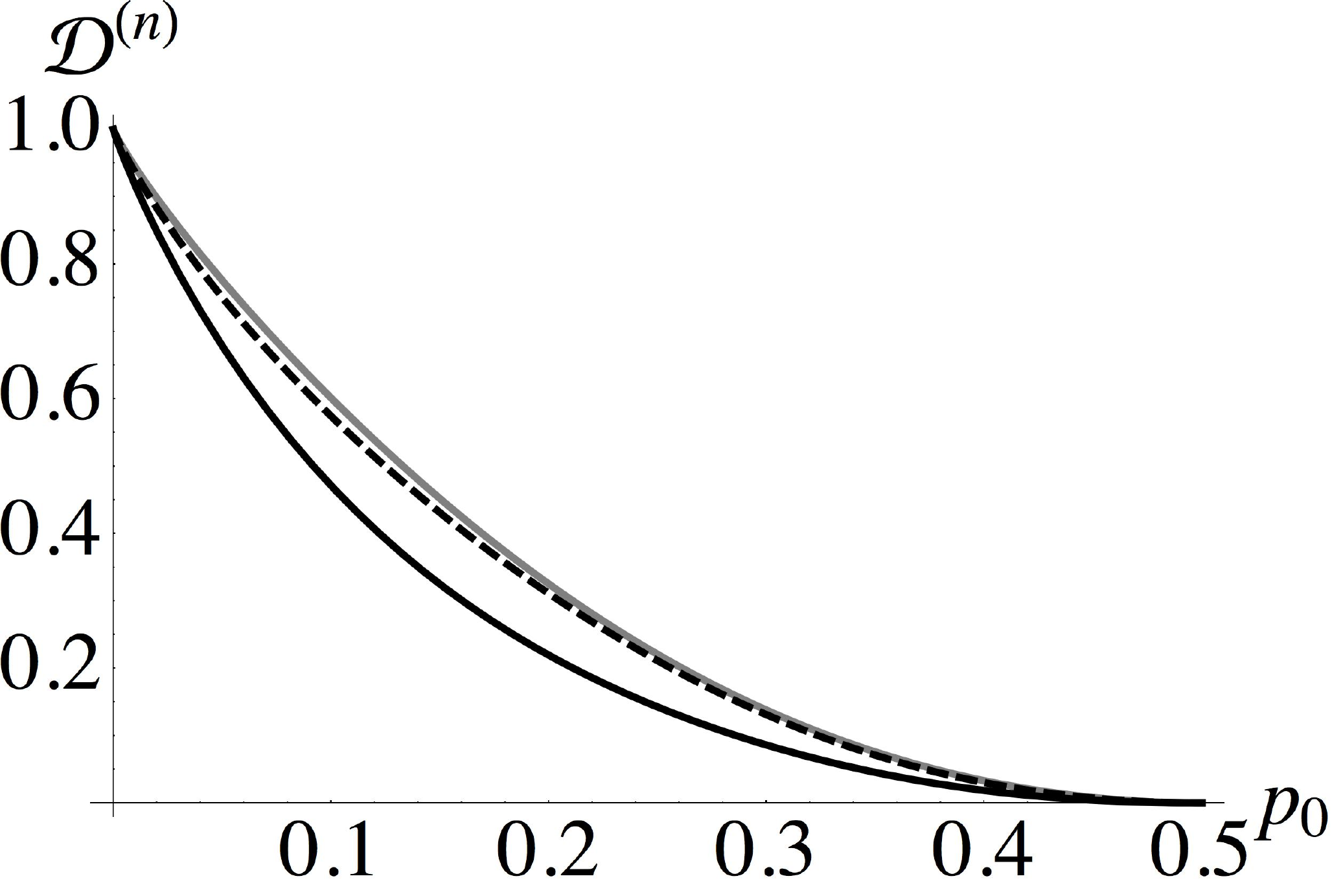}~~~~~~~~~~\includegraphics[width=5cm]{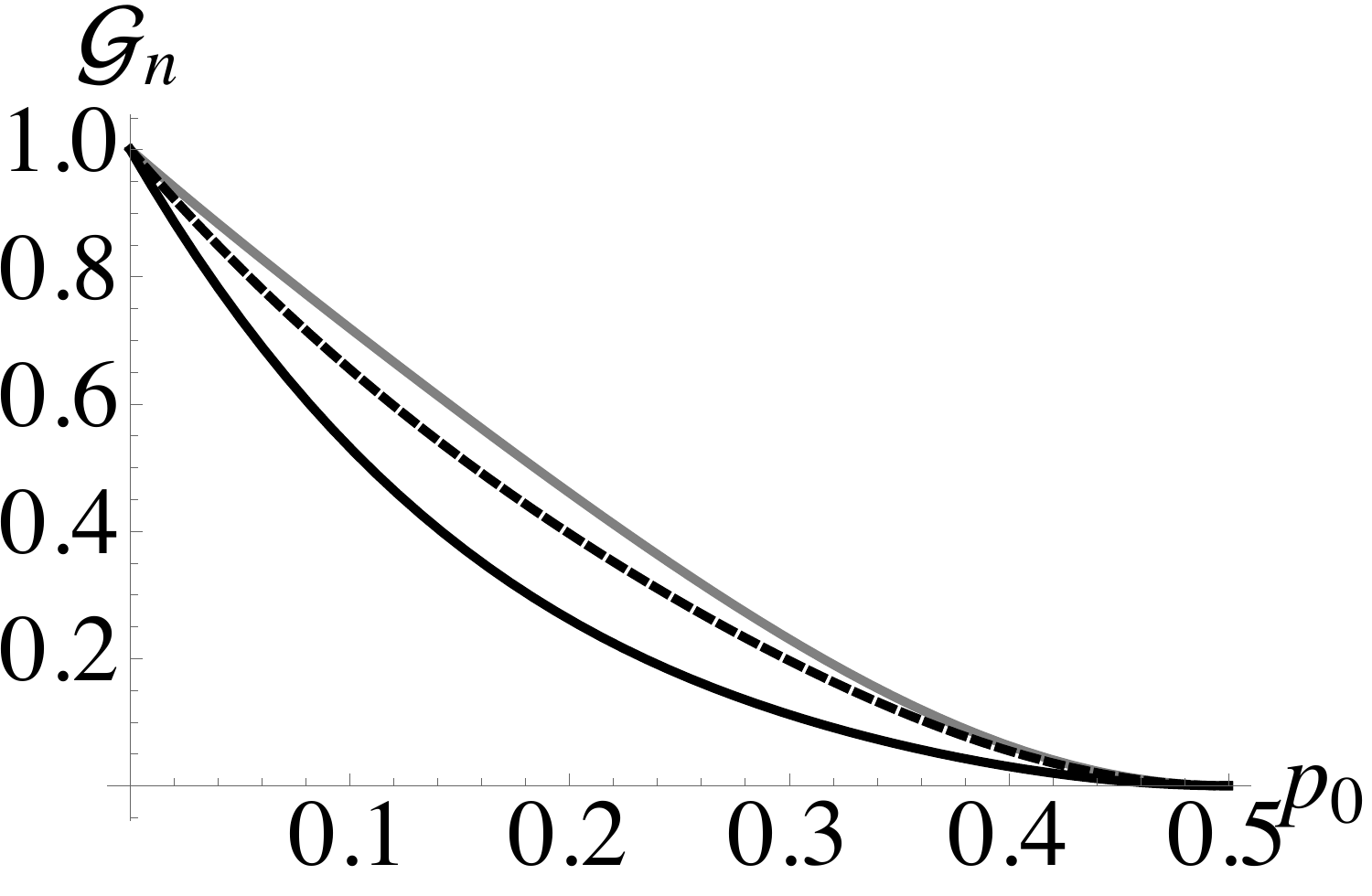}
\vspace*{8pt}
\caption{ (a) Genuine quantum correlations as measured by ${\cal D}^{(n)}$ for the case of three (gray), four (dashed black) and six (solid black) qubits. The state considered is the one given in Eq.~(\ref{rho}). For $p_0=1/2$ the state becomes a product one and any correlation disappears. (b) As for panel (a) calculating the global discord. Both plots are symmetric under the exchange $p_0 \leftrightarrow p_1$.  }
\label{fig1}
\end{center}
\end{figure}

For Eq.~(\ref{rho}) we can also efficiently calculate analytically the global discord. The first simplification is noting that the final term in Eq.~(\ref{GQD}) corresponds to a relative entropy for each individual qubit, which due to the fully symmetric nature of the state we can fully determine this term by calculating it for any single qubit and taking $n$ times this. The most difficult part comes from calculating the relative entropy for the total state given by the first term in Eq.~(\ref{GQD}), since for an arbitrary $n$ qubit state the minimization requires $2n$ angles. However, since the state is fully symmetric it is clear that only a single angle is required, as it is the same applied to each qubit. For the state at hand we find the angle required to minimize Eq.~(\ref{GQD}) is $\theta=\frac{\pi}{2}$. Finally we find the GD for Eq.~(\ref{rho})
\begin{equation}
\label{GDanalytic}
\mathcal{G}_n=p_0^n \log_2 p_0^n+p_1^n \log_2 p_1^n-\left(p_0^n+p_1^n\right) \log_2 \left(\frac{1}{2} \left[p_0^n+p_1^n\right]\right),
\end{equation}
In Fig.~\ref{fig1} (b) we show the behavior of ${\mathcal G}_n$ as a function of the state parameter $p_0$ for different values of $n$.

\section{Multipartite correlations in open systems}
As a second, physically relevant application, we assess
 the multipartite measures and indicators in open systems. In particular we study the broadly applicable amplitude and phase damping channels applied to the generalized $n$-qubit $GHZ$ multipartite entangled states. The amplitude damping (AD) channel describes the probability of losing an excitation to the surrounding environment. By modeling the environment, $E$, as a qubit interacting with the system qubit, $S$, the action of the amplitude damping channel is
\begin{eqnarray*}
&\ket{0_S 0_E}&\to\ket{0_S 0_E},\\
&\ket{1_S 0_E}&\to \sqrt{1-\lambda} \ket{1_S 0_E}+ \sqrt{\lambda} \ket{0_S 1_E}.
\end{eqnarray*}
The phase damping (PD) channel acts in a similar manner, the only difference being this affects only the coherence present in the system and leaves the populations (i.e. the energy) unchanged. The phase damping channel acts as
\begin{eqnarray*}
&\ket{0_S 0_E}&\to\ket{0_S 0_E},\\
&\ket{1_S 0_E}&\to \sqrt{1-\gamma} \ket{1_S 0_E}+ \sqrt{\gamma} \ket{1_S 1_E}.
\end{eqnarray*}
We apply these channels locally to each qubit of the multipartite states (assuming the same damping rate for all qubits). By then tracing over the all environment degrees of freedom we determine the density matrices for the various $n$-qubit noisy states. Starting from the generalized $n$-qubit $GHZ$ state
\begin{equation}
\ket{\psi_{GHZ}}= \alpha_1 \ket{0_1\dots0_n} +\alpha_2 \ket{1_1\dots1_n},~~~~~\alpha_1\in[0,\frac{1}{\sqrt{2}}],~\alpha_2=\sqrt{1-\alpha_1^2},
\end{equation}
we find the corresponding multipartite noisy states take the form
\begin{eqnarray}
\label{GHZAD}
\nonumber
\varrho_{GHZ}^{AD}&=&(\alpha_1^2+\alpha_2^2 \lambda^n) \ket{0_1\dots0_n}\bra{0_1\dots0_n}+\alpha_2^2(1-\lambda)^n \ket{1_1\dots1_n}\bra{1_1\dots1_n} \\
                          &+&\alpha_1\alpha_2(1-\lambda)^\frac{n}{2} \left( \ket{0_1\dots0_n}\bra{1_1\dots1_n} + \ket{1_1\dots1_n}\bra{0_1\dots0_n} \right) \\
                           &+& \alpha_2^2 \sum^{n-1}_{k=1} (1-\lambda)^k \lambda^{n-k} \mathcal{I}_k \nonumber
\end{eqnarray}

\begin{eqnarray}
\label{GHZPD}
\varrho_{GHZ}^{PD} &=& \alpha_1^2 \ket{0_1\dots0_n}\bra{0_1\dots0_n}+\alpha_2^2 \ket{1_1\dots1_n}\bra{1_1\dots1_n} \\
               &+& \alpha_1\alpha_2(1-\gamma)^\frac{n}{2} \left( \ket{0_1\dots0_n}\bra{1_1\dots1_n} + \ket{1_1\dots1_n}\bra{0_1\dots0_n} \right) \nonumber
\end{eqnarray}

\subsection{Genuine correlations}
The same derivation described in Sec. \ref{sec:gen} can be applied to the case of a multipartite GHZ state under amplitude damping.
In fact, $\varrho_{nGHZ}^{AD}$ is both $\mathcal{T}$- and $\mathcal{P}^{(k)}$-symmetric and all the considerations made before are valid. The optimal value of $\theta$, however, will not be generally equal to $\pi/4$ any more. 

As for the case of $\varrho_{GHZ}^{PD}$, the solution is even simpler, given that the state is rank $2$, and, irrespective of the cut considered, it is always possible to purify it adding an external qubit as ancilla~\cite{epl,r2}. Then, the Koashi-Winter formula can be used to find an analytical expression for the multipartite discord~\cite{kw}. For instance, in the tripartite case, we have
\begin{equation}
{\cal D}^{(3)}(\varrho_{ijk})=\min\{S(\varrho_{jk})-S(\varrho)+{\cal E}(\varrho_{i,a}),S(\varrho_{i})-S(\varrho)+{\cal E}(\varrho_{jk,a}) \}
\end{equation}
where $a$ denotes the ancilla and ${\cal E}(.)$ is the entanglement of formation, which can be calculated analytically, since in the $jk$ part only two levels are populated.

\begin{figure}[t]
\begin{center}
(a) \hskip5cm (b)
\includegraphics[width=6cm]{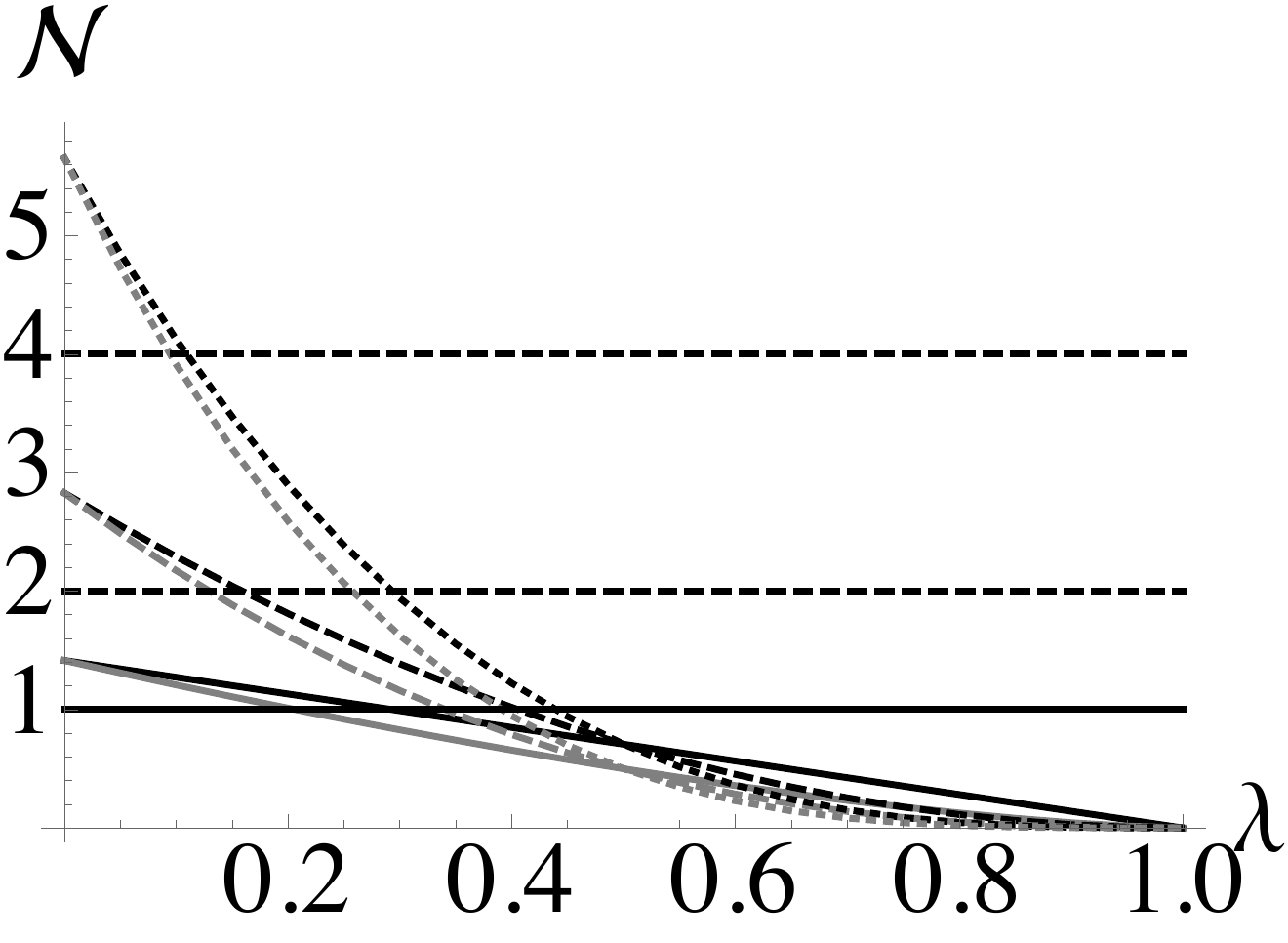}~~~~~~~~~~\includegraphics[width=6cm]{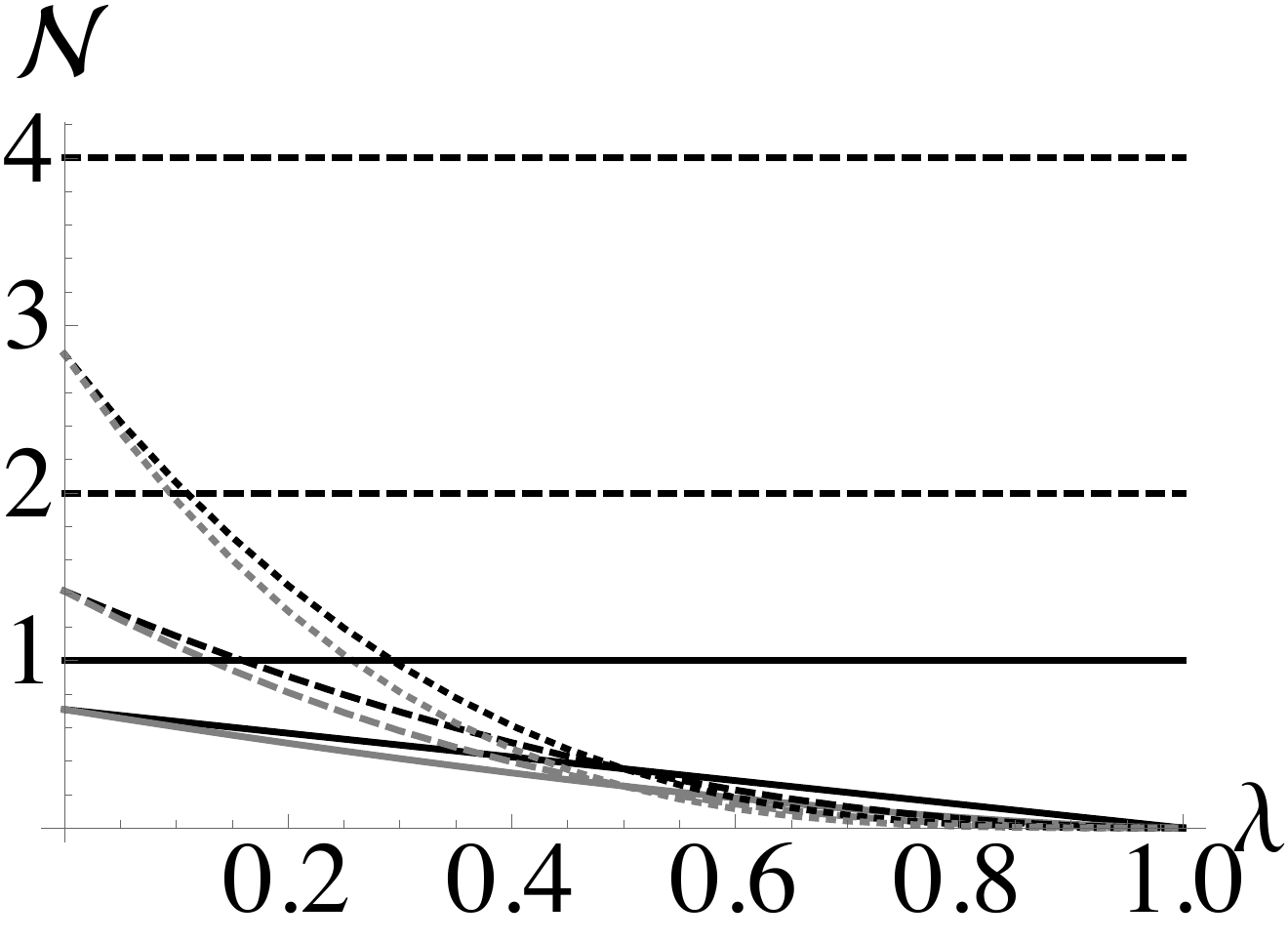}
\vspace*{8pt}
\caption{ (a) Nonlocality for an initially ideal GHZ state. Curves for $n=$ 2, 3 (solid), 4, 5 (dashed), and 6, 7 (dotted) qubits from bottom to top. Lowest horizontal line at $\mathcal{N}=1$ is the local hidden variable bound, violation of this implies our state is in some way non-local. The other dashed horizontal lines are the bounds for $1:n-1$ separability for $4$ or 5 and 6 or 7 qubits respectively. (b) As for panel (a) starting from a weighted $GHZ$ state with  $\alpha_1=\frac{\sqrt{2+\sqrt{3}}}{2}$.}
\label{fig3}
\end{center}
\end{figure}

\subsection{Nonlocality}
Determining the non-locality is again a difficult task because in order to search for a violation of the inequality we are required to maximize over the two chosen measurement settings. This is a state-dependent issue, meaning that for a given state we need to be careful precisely what measurement settings we choose, and regardless, as with the calculation of discord, there is an optimization required. Each of the different elements entering the Svetlichny inequalities Eq.~(\ref{SI}) corresponds to an $n$-partite correlation function determined by applying local rotations to each qubit. For the damped $GHZ$ state Eq.~(\ref{GHZAD}) we apply the local rotation $\mathcal{R}(\theta_i^{q_i})=\cos(\theta_i^{q_i}) \hat\sigma_x+\sin(\theta_i^{q_i})\hat\sigma_y$ to each of the $n$ qubits with $q_i=1$ or 2 being the two possible measurement settings. We find the correlation function takes a simple form
\begin{eqnarray}
\label{corr}
\mathcal{C}(\theta_1^{q_1} \dots \theta_n^{q_n})& =& \mathrm{Tr}\left[\mathcal{R}(\theta_1^{q_1} \dots \theta_n^{q_n}) \varrho_{GHZ}^{AD}\right]  \\
                                                                      &=&2 (1-\lambda)^\frac{n}{2} \alpha_1 \alpha_2 \cos\left(\theta_1^{q_1} +\dots + \theta_n^{q_n}\right), \nonumber
\end{eqnarray}
with $q_i=1$ or 2. All that remains is to iteratively determine the corresponding Svetlichny polynomial for a given $n$ to study the non-locality. For clarity, let us explicitly determine this for $n=2$. We have then from Eq.~(\ref{SI})
\begin{eqnarray*}
\mathcal{N}_2=m_2&=&\frac{1}{2}o_{1}(o_2+O_2)+\frac{1}{2}O_{1}(o_2-O_2)\\
       &=&\frac{1}{2}\left( o_1 o_2 +o_1 O_2 + O_1o_2 - O_1 O_2  \right),
\end{eqnarray*}
with each term inside the brackets corresponding to a correlation function. Using Eq.~(\ref{corr}) with $n=2$ we have
\begin{eqnarray*}
\mathcal{N}_2&=&\frac{1}{2}\left(  \mathcal{C}(\theta_1^1,\theta_2^1) + \mathcal{C}(\theta_1^1,\theta_2^2) + \mathcal{C}(\theta_1^2,\theta_2^1) - \mathcal{C}(\theta_1^2,\theta_2^2)  \right)\\
                       &=&(1-\lambda) \alpha_1 \alpha_2 \left(  \cos\left(\theta_1^{1} + \theta_2^{1}\right) + \cos\left(\theta_1^{1} + \theta_2^{2}\right) + \cos\left(\theta_1^{2} + \theta_2^{1}\right) - \cos\left(\theta_1^{2} + \theta_2^{2}\right) \right)
\end{eqnarray*}
which is the well known CHSH inequality for a 2 qubit generalized Bell state. Larger $n$ then follows iteratively applying Eq.~(\ref{SI}) in the same fashion, and we find there are $2^n$ distinct correlation functions appearing in the inequality. It is interesting to notice that if we calculate the same quantities for the PD affected $GHZ$ state, Eq.~(\ref{GHZPD}), we find precisely the same result. This means that the non-locality is not sensitive the type of noise the state is undergoing. In Fig.~\ref{fig3} we show the non-locality for a $n$-qubit $GHZ$ state undergoing either AD or PD. In panel (a) $\alpha_1=\frac{1}{\sqrt{2}}$, i.e. we start from a perfect $GHZ$ state. A few interesting features to note, the larger the system the more noise it can withstand before it no longer violates the classical bound of $1$ (lowest solid horizontal line). The dashed horizontal lines in the plots at $2$ and $4$, correspond to the bounds for $1:(n-1)$ separability for $4$ or $5$ and $6$ or $7$ qubits respectively. Values below these lines for a general state means we would not be certain the violation of Eq.~(\ref{SI}) was due to full $n$-partite or $n-1$-partite correlations, however given the initial form of the state we know in this special instance that our violation arises due to full $n$-partite correlations. In panel (b) we take $\alpha_1=\frac{\sqrt{2+\sqrt{3}}}{2}$, this state is initially much less correlated. A qualitatively similar behavior holds, however, we never see large violations of the inequalities and for very small systems $n=2,3$ there is no violation at all.

\section{Conclusions}
Even if calculating correlations in multipartite states is a hard and generally unsolved problem, the existence of symmetries greatly simplifies the task. In this work we have addressed the quantitative analysis of quantum correlations in some broadly applicable, and physically relevant classes of states. Inspired by maximal work extraction protocol from a cyclic transformation (ergotropy), we have shown how to calculate genuine and global discord under eigenstate swap. Using the same symmetry considerations, we have also studied the behavior of multipartite nonlocality, together with discord, in the presence of local noise. Beyond the quantification of the correlations in this instance, we also noted a stark difference between the non-locality and discord; while the discord was sensitive to the type of lossy channel applied, the nonlocality was unable to discriminate which noisy process the state was undergoing. It remains an open question how the analysis extends to higher dimensional systems, where the calculation of such figures of merit becomes even more involved.

\section*{Acknowledgments}
SC is funded through the EU Collaborative Project TherMiQ (Grant Agreement 618074), GLG acknowledges financial support from Compagnia di San Paolo and from the EU Collaborative project QuProCS (Grant Agreement 641277).


\end{document}